%% file: Pair-01-16.tex
\documentclass[12pt]{article}

\usepackage{setspace}
\usepackage[left=1in,right=1in,top=1in,bottom=1.5in]{geometry}

\usepackage[authoryear]{natbib}

\usepackage{subfigure}
\usepackage{amsfonts}
\usepackage{amsmath}
\usepackage{bm}
\usepackage{amsthm}
\usepackage{enumerate}

\usepackage{caption}
\usepackage{hyperref}

\usepackage{xcolor}
\usepackage[normalem]{ulem}

\usepackage{enumitem}
\newcommand{\subscript}[2]{$#1 #2$}

\usepackage{array,epsfig,fancyheadings,rotating}
\usepackage[]{hyperref}  
\usepackage{sectsty, secdot}
\sectionfont{\fontsize{12}{14pt plus.8pt minus .6pt}\selectfont}
\renewcommand{\theequation}{\thesection\arabic{equation}}
\subsectionfont{\fontsize{12}{14pt plus.8pt minus .6pt}\selectfont}

\hypersetup{colorlinks=true, linkcolor=blue, citecolor=black, urlcolor=blue}

\newtheorem{theorem}{Theorem}
\newtheorem{lemma}{Lemma}
\newtheorem{corollary}{Corollary}
\newtheorem{proposition}{Proposition}
\theoremstyle{definition}

\newtheorem{remark}{Remark}

\newcommand{\tcorr}{\text{Corr}}
\newcommand{\hcorr}{\widehat{\tcorr}}

\newcommand{\RNum}[1]{\uppercase\expandafter{\romannumeral #1\relax}}

\newcommand{\PP}[1]{\mathrm{P}\left(#1\right)}

\newcommand{\Aa}{A_{\alpha}}
\newcommand{\Ba}{B_{\alpha}}
\newcommand{\ea}{\eta_{\alpha}}
\newcommand{\ainI}{\alpha\in I}
\newcommand{\lambn}{\lambda_{p,n}}

\newcommand{\rij}{\rho_{i,j}}
\newcommand{\hrij}{\hat{\rho}_{ij}}

\newcommand{\nnpn}{p^{-4/(n-2)}}
\newcommand{\aapn}{a_{p,n}}
\newcommand{\bbpn}{b_{p,n}}
\newcommand{\ccpn}{c_{p,n}}
\newcommand{\detnp}{\delta_{p,n}}

\newcommand{\spn}{S^2_{pn}}

\newcommand{\Wpn}{W_{pn}}
\newcommand{\Rpn}{R^2_{pn}}
\newcommand{\Tpn}{T_{pn}}

\newcommand{\Betan}{B(\frac{1}{2}, \frac{n-2}{2})}

\newcommand{\hbeta}{\hat{\beta}}

\newcommand{\veps}{\varepsilon}

\newcommand{\rar}{\rightarrow}
\newcommand{\limn}{\lim_{n \rightarrow \infty}}

\newcommand{\infn}{n \rightarrow \infty}
\newcommand{\infp}{p \rightarrow \infty}

\newcommand{\calc}{\mathcal{C}}
\newcommand{\calg}{\mathcal{G}}
\newcommand{\calm}{\mathcal{M}}
\newcommand{\caln}{\mathcal{N}}

\newcommand{\bx}{\mathbf{x}}
\newcommand{\by}{\mathbf{y}}
\newcommand{\bz}{\mathbf{z}}

\newcommand{\bw}{\mathbf{w}}
\newcommand{\bveps}{\boldsymbol{\veps}}
\newcommand{\bbeta}{\boldsymbol{\beta}}

\begin{document}

\fontsize{12}{14pt plus.8pt minus .6pt}\selectfont \vspace{0.8pc}
\centerline{\large\bf Penalized linear regression with  }
\vspace{2pt} \centerline{\large\bf high-dimensional pairwise screening}
\vspace{.4cm} \centerline{Siliang Gong, Kai Zhang and Yufeng Liu} \vspace{.4cm} \centerline{\it
	University of Pennsylvania}
\centerline{\it and The University of North Carolina at Chapel Hill} \vspace{.55cm} \fontsize{9}{11.5pt plus.8pt minus
	.6pt}\selectfont

\begin{quotation}
	\noindent {\it Abstract:}
	{In variable selection, most existing screening methods focus on marginal effects and ignore dependence between covariates. To improve the performance of selection, we incorporate pairwise effects in covariates for screening and penalization. We achieve this by studying the asymptotic distribution of the maximal absolute pairwise sample correlation among independent covariates. The novelty of the theory is in that the convergence is with respect to the dimensionality $p$, and is uniform with respect to the sample size $n$. Moreover, we obtain an upper bound for the maximal pairwise R squared when regressing the response onto two different covariates. Based on these extreme value results, we propose a screening procedure to detect covariates pairs that are potentially correlated and associated with the response. We further combine the pairwise screening with Sure Independence Screening \citep{fan08sis} and develop a new regularized variable selection procedure.  Numerical studies show that our method is very competitive in terms of both prediction accuracy and variable selection accuracy.}\\
	
	\vspace{1pt}
	\noindent {\it Key words and phrases:Pairwise Screening, Penalized Regression, Sure Independence Screening, Variable Selection}

	\par
\end{quotation}\par

\renewcommand{\theequation}{\thesection.\arabic{equation}}

\fontsize{12}{14pt plus.8pt minus .6pt}\selectfont

\section{Introduction}\label{sec:pairintro}

In the era of big data, high dimensional problems are of interest in many scientific fields, where the number of variables may be comparable to or even much larger than the sample size. For example, in genetic studies, one often has tens of thousands of genes in the microarray datasets with only a few hundreds of patients; in neuroscience, fMRI images may contain millions of voxels. 

In recent years, much research effort has been devoted to dealing with high dimensional data analysis. Among those methods developed, penalized least squares plays an important role. In particular, one of the most well-known method is the LASSO proposed by \cite{tib96}, which is the solution to the following penalized problem
\begin{equation}\label{eq:penlike}
\min_{\bbeta\in\mathbb{R}^p} \|\by-X\bbeta\|_2^2+\lambda P(\bbeta),
\end{equation} 
where $\lambda P(\bbeta)=\lambda \sum_{j=1}^p |\beta_j|$ is the $l_1$ penalty. 
\cite{tib96} showed that the LASSO leads to a sparse estimator that shrinks the OLS solution and sets some of the estimated coefficients to exact zero. Despite with good theoretical properties and practical performance, the LASSO has two major drawbacks: firstly, due to the shrinkage nature, LASSO may over-shrink the estimates and cause significant bias; secondly, if there is a group of variables that are highly correlated, LASSO tends to select only one of them. To address these issues, \cite{zou05elnet} introduced the elastic net method, using $\lambda_1\|\bbeta\|_1+\lambda_2\|\bbeta\|^2_2$ as the regularization term in \eqref{eq:penlike} and thus encouraging a grouping effect. Besides the elastic net, various penalized variable selection methods have been proposed as extensions to LASSO, including the Dantzig selector \citep{candes07}, the smoothly clipped absolute deviation (SCAD) penalty \citep{fan01noncon}, among many others. See \cite{friedman01elements} and \cite{fan10selective} for a comprehensive overview.

For high dimensional variable selection, it is crucial to account for the dependency structure of covariates. Such structure information not only improves the accuracy of selection, but also have practical meanings. For instance, in gene expression data, genes usually function as biological pathways instead of working independently. Classical penalized variable selection methods, however, usually do not explicitly take into account the relationships between covariates. To address this problem, \cite{yuan06} proposed the group LASSO method, which takes advantage of the grouping information among the covariates. Extensions of group lasso include, but are not limited to \cite{breheny15group}. Other methods use the structure information as predictor graph (see \cite{li08network, pan10, zhu13sim, yu2016} among others for reference). 

A common assumption for the methods mentioned above is that the underlying predictor graph is given, which may not hold in practice. When the prior information is not available, the idea of clustering can be incorporated to improve regression performance. Specifically, \cite{park07} proposed to perform hierarchical clustering on the covariates first and take the cluster average as new predictors for regression. There are also methods using supervised clustering to encourage highly correlated pairs of covariates to be included or excluded simultaneously \citep{bondell08,sharma13}. Similarly, another type of methods aims to make correlated covariates have similar regression coefficients \citep{she10sparse}. Nevertheless, a large sample correlation between two variables does not necessarily indicate that they are dependent in the population sense. When the dimensionality continues to increase, the maximal pairwise correlation among $p$ independent covariates can be close to 1 \citep{fan10selective}. Therefore, it is important to identify covariates that are truly correlated and incorporate such information into variable selection procedures. 

In this paper, we study the limiting behavior of the maximal absolute pairwise sample correlation among covariates when they are independent Gaussian random variables. Different from existing work, we investigate the limiting distribution as the dimensionality $p$ diverges. Therefore, the proposed asymptotic results potentially can be applied to datasets with arbitrarily large  dimensionality. We further discuss the extreme behavior of the maximal absolute Spearman's rho statistic for covariates with general distributions. On the other hand, we obtain the upper bound of maximal pairwise R squared when regressing the response onto pairs of covariates. With the extreme value results, we formulate a screening procedure to identify covariates pairs that are potentially dependent and associated with the response. We further combine the pairwise screening with the Sure Independence Screening (SIS) \citep{fan08sis} and propose a novel penalized variable selection method. More specifically, we assign different penalties to each individual covariate according to the screening results. Numerical experiments show that the performance of our proposed method is competitive compared with existing approaches in terms of both variable selection and prediction accuracy. 

The remainder of this paper is organized as follows: We first investigate the limiting distribution of the maximal pairwise sample correlation among covariates in Section \ref{sec:extreme}. We also show that our asymptotic results cover that of \cite{cai12phase} as a special case. Then we propose an upper bound for the maximal pairwise R squared in Section \ref{sec:rsquare}. In Section \ref{sec:pen} we formulate our proposed variable selection approach as a penalized maximum likelihood problem, and discuss potential extensions of our method in Section \ref{sec:extend}. Theoretical properties are discussed in Section \ref{sec:theo}. We show with simulated experiments as well as two real datasets in Section \ref{sec:pairns} that the proposed method has improved performance when important variables are highly correlated. Finally, we conclude this paper and discuss possible future work in Section \ref{sec:pairsum}. Proofs of the theoretical results are provided in the Appendix.

\section{Pair Screening for covariates}\label{sec:pairtest}

Suppose we have the following linear model
\begin{equation}\label{eq:pairlm}
\by = X\bbeta +\bveps,
\end{equation}
where $\by=(y_1, y_2, \cdots, y_n)^T$ is the response vector, $X = (\bx_1, \bx_2, \cdots, \bx_p)$ is an $n\times p$ design matrix with $\bx_j$ being $n$ independent and identical observations from the covariate $X_j$. We assume that the covariate vector $\bx = (X_1, X_2, \cdots, X_p)^T$ has a multivariate distribution with unknown covariance matrix $\Sigma$, and $\bveps = (\veps_1, \veps_2, \cdots, \veps_n)^T$ is a vector of i.i.d. random variables with mean 0 and standard deviation $\sigma$, {and is independent of the covariate vector $\bx$}. 

For the linear model \eqref{eq:pairlm}, variable selection methods aim to identify the non-zero components of $\bbeta$, in other words, the important variables among all candidate predictors. Particularly, if two covariates have a large pairwise correlation, we may want to include or exclude these two variables simultaneously when conducting variable selection. However, the sample correlation can be spurious, especially when the number of covariates $p$ is relatively large. Therefore, it is important to identify covariates that are truly correlated. In other words, we need to find a threshold for the pairwise sample correlation among covariates to screen covariates pairs. In the following subsection, we will discuss in details the asymptotic results that generate the screening rule.

\subsection{Extreme laws of pairwise sample correlation among covariates}\label{sec:extreme}

We propose to choose a bound based on the extreme laws of the pairwise sample correlation when the $p$ covariates are independent. Our investigations are under two settings: (a) the covariates are normally distributed; (b) the covariates are non-Gaussian random variables.  

\subsubsection{Gaussian covariates}

It has been recently studied that the maximal absolute Pearson sample correlation between $p$ i.i.d. Gaussian covariates and an independent response has a Gumble-type limiting distribution as $p$ goes to infinity \citep{zhang15pack}. Motivated by \cite{zhang15pack}'s work, we find that the maximal absolute pairwise sample correlation among $p$ independent covariates also has a limiting distribution, as stated in the following theorem:

\begin{theorem}\label{thm:wdist}
	Suppose $X_1, X_2, \cdots, X_p$ are $p$ independent	Gaussian variables and we observe $n$ independent samples from each of $X_j$'s. Let $\Wpn = \max_{1\le i<j\le p}|\rij|$, where $\rij = \hcorr(X_i, X_j)$ is the Pearson sample correlation between $X_i$ and $X_j$. Then as $\infp$, 
	\begin{equation}\label{eq:pairdist}
	\lim_{p\rightarrow \infty}|P(\frac{\Wpn^2-\aapn}{\bbpn}\le x) - I(x\le \frac{n-2}{2}) \exp\big\{-\frac{1}{2}\big(1-\frac{2}{n-2}x\big)^{\frac{n-2}{2}}\big\}-I(x>\frac{n-2}{2})|=0,
	\end{equation}
	which is uniformly for any $n\ge 3$. Here  
	$\aapn = 1-\nnpn\ccpn$,
	$\bbpn = \frac{2}{n-2}\nnpn\ccpn$, and  
	$\ccpn=\big(\frac{n-2}{2}\Betan\sqrt{1-\nnpn}\big)^{2/(n-2)}$
	are the normalizing constants.
\end{theorem}

In random matrix theory, $\Wpn$ is also known as the coherence when the design matrix $X$ is random. Specifically, the coherence is defined as the largest magnitude of the off-diagonal entries of the sample correlation matrix associated with a random matrix. The limiting behavior of the coherence has been well studied when the sample size $n$ goes to infinity. For example, \cite{cai11limiting} studied the asymptotic distribution under certain regularity conditions with application to the testing of covariance matrix. \cite{cai12phase} further obtained the limiting laws of the coherence for different divergence rate of $p$ with respect to $n$ and summarized the results as phase transition phenomena. We can show that our result unifies the convergence in terms of the sample size, and covers \cite{cai12phase}'s as special cases, described in the following corollary.

\begin{corollary}\label{col:comp}
	Let $\Wpn$ be defined as in Theorem \ref{thm:wdist}, where we still assume $X_j$'s are independent normal random variables. Let $\Tpn = \log(1-\Wpn^2)$.
	\begin{enumerate}
		\item [(a)]
		(\textbf{Sub-Exponential Case})
		Suppose $p = p_n\rightarrow\infty$ as $\infn$ and $(\log p)/n\rightarrow 0$, then as $\infn$,
		\begin{equation*}
		\PP{n\Tpn+4\log p-\log\log p\le x}\rar 1-e^{-\frac{1}{\sqrt{8\pi}}e^{x/2}}.
		\end{equation*}
		\item [(b)]
		(\textbf{Exponential Case}) 
		Suppose $p = p_n$ satisfies $(\log p)/n \rightarrow \beta\in (0, \infty)$ as $\infn$. Then as $\infn$,
		\begin{equation*}
		\PP{n\Tpn+4\log p-\log\log p\le x} \rar 
		1-\exp\big\{K(\beta)e^{(x+8\beta)/2}\big\},
		\end{equation*}
		where $K(\beta) = \big(\frac{\beta}{2\pi(1-4e^{-4\beta})}\big)^{1/2}$.
		\item [(c)]
		(\textbf{Super-Exponential Case})
		Suppose $p = p_n$ satisfies $(\log p)/n\rar \infty$  as $\infn$. Then as $\infn$,
		\begin{equation*}
		\PP{n\Tpn + \frac{4n}{n-2}\log p-\log n \le x}
		\rar 1-e^{-\frac{1}{\sqrt{2\pi}}e^{x/2}}.
		\end{equation*}
	\end{enumerate}
\end{corollary}

Compared with previous work, our asymptotic distribution is novel in two aspects. First, the convergence in Theorem \ref{thm:wdist} is with respect to $p$ instead of $n$, making it applicable to high dimensional data, or even ultrahigh dimensional problems. Moreover, the convergence result we have discovered is uniform for any $n\ge 3$, thus finite sample performance is guaranteed. 

\subsubsection{Non-Gaussian covariates}

When the covariates are non-Gaussian random variables, it is more desirable to choose a distribution-free statistic for the screening rule. Therefore, instead of using the Pearson's sample correlation, we study the extreme behavior of the Spearman's rho statistic \citep{spearman1904}. Recall that $\bx_j = (X_{1j}, X_{2j}, \cdots, X_{nj})^T$ are $n$ i.i.d. observations from the covariate $X_j$. Let $Q_{ni}^{j}$ and $Q_{ni}^k$ be the ranks of $X_{ij}$ and $X_{ik}$ in $\{X_{1j},\cdots, X_{nj} \}$ and $\{X_{1k}, \cdots, X_{nj}\}$ respectively. Then the Spearman's rho is defined as 
\begin{equation}
\rho_{ij} = \frac{\sum_{i=1}^n (Q_{ni}^j-\bar{Q}_n^j)(Q_{ni}^k - \bar{Q}_n^k)}{\sqrt{\sum_{i=1}^n (Q_{ni}^j-\bar{Q}_n^j)^2\sum_{i=1}^n(Q_{ni}^k - \bar{Q}_n^k)^2}},
\end{equation}
where $\bar{Q}_n^j = \bar{Q}_n^k = \frac{n+1}{2}$.

Similar to the normal setting, we are particularly interested in the limiting distribution of $\spn = \max_{1\le i < j\le p} \rho^2_{ij}$ when the covariates are all independent, which has been studied in \cite{han2014}. The following proposition states that as $n$ increases, $\spn$ converges to a Gumble type distribution. 

\begin{proposition}\label{thm:spear}
	Suppose that $X_1, \cdots, X_p$ are independent and identically distributed random variables, and we have $n$ independent samples for each of the covariates. Let $\spn = \max_{1\le i < j\le p} \rho^2_{ij}$ be the squares of the maximal pairwise Spearman's rho statistics,  then for $\log p =o(n^{1/3})$, we have
	\begin{equation}
	\limn|\PP{(n-1)\spn-4\log p+\log\log p \le x} -\exp\big\{-(8\pi)^{-1/2}\exp(-x/2)\big\} |=0.
	\end{equation}
\end{proposition} 

Theorem \ref{thm:wdist} and Proposition \ref{thm:spear} characterize the magnitude of the maximal pairwise correlation and Spearman's rho statistic respectively when the covariates are independent. {Suppose a pair of covariates, say $X_1$ and $X_2$, have a absolute sample correlation greater than the $95\%$ quantile of the distribution given in Theorem \ref{thm:wdist} or Proposition \ref{thm:spear}, then they tend to be marginally dependent. Since we are only interested in pairs of truly important covariates,} we further investigate the extreme behavior of the maximal pairwise R squared under the null model, i.e., $\beta_j$'s are all equal to zero.

\subsection{R squared screening for pairs of covariates}\label{sec:rsquare}

With the asymptotic distributions introduced in the previous subsections, we can identify covariates pairs that are potentially dependent. However, such screening does not take into account the association between the covariates and the response. It is possible that an important variable has a large sample correlation with unimportant ones; or two highly correlated covariates are both unrelated to the response. To address such an issue, we introduce another screening procedure based on the R squared from regressing the response $Y$ onto the pairs of covariates.

Consider the linear regression where we regress $Y$ onto a pair of covariates $X_i$ and $X_j$
with $i\ne j$, we can obtain the corresponding R squared $R^2_{ij}$. Under the model setting \eqref{eq:pairlm}, when all the coefficients are zeros, the maximal pairwise R squared $\displaystyle\max_{1\le i <j \le p} R^2_{ij}$ cannot be too large. In fact, there exists an asymptotic bound for $\displaystyle\max_{1\le i <j \le p} R^2_{ij}$, as described in the following theorem. 

\begin{theorem}\label{thm:rsquare}
Let $\Rpn = \max_{1\le i<j\le p}R^2_{ij}$, where $R^2_{ij}$ is the pairwise $R$ squares from regressing $Y$ onto $X_i$ and $X_j$ where $i\ne j$. Suppose that $X_1,  \cdots, X_p$ and $Y$ are from the model setting \ref{eq:pairlm} and we further assume that $Y$ is a normally distributed. Then when $\beta_j$'s are all zeros, we have for any fixed $n\ge 4$, $\delta>0$, as $p\rightarrow\infty$, $P(\Rpn\ge 1-p^{-(4+\delta)/(n-3)})= O(p^{-\delta/2}) \rightarrow 0$.
\end{theorem}

With the bound given by Theorem \ref{thm:rsquare}, we can design a screening rule to find pairs of covariates that are potentially associated with the response. In Section \ref{sec:penalty}, we introduce how to make use of the theoretical results to benefit variable selection.

\section{Penalized variable selection using pairwise screening}
\label{sec:penalty}

In this section, we propose a pairwise screening procedure that takes advantages of the asymptotic results in Section \ref{sec:pairtest}. We further establish a new penalization algorithm for variable selection.

\subsection{Screening-based penalization}\label{sec:pen}
Given the limiting distribution of the maximal pairwise sample correlation described in Section \ref{sec:pairtest}, we propose the following screening rule to identify covariates pairs that are potentially correlated and related to the response:
\begin{equation}\label{eq:screen}
\calg = \{(i,j): i < j, |\hcorr(X_i, X_j)|\ge a \text{ and } R^2_{ij}\ge r_0 \},
\end{equation}
where $a$ is the $100(1-\alpha)\%$ quantile of the distribution given in Theorem \ref{thm:wdist} (for Gaussian covariates) or Proposition \ref{thm:spear} (for non-Gaussian covariates), and $r_0 = 1-p^{-(4+\delta)/(n-3)}$. Note that the values of $\alpha$ and $\delta$ can affect the size of $\calg$. The larger $\alpha$ and $\delta$ are, there are fewer pairs included in $\calg$. In practice, we suggest to take $\alpha = 0.05$ and $\delta=0.1$.

The group definition in \eqref{eq:screen} is a screening procedure with respect to covariates pairs. The idea of screening is prevalent for high dimensional data analysis. In particular, for penalized variable selection methods, increasing dimensionality makes it more difficult to capture the inherent sparsity structure. Therefore, dimension reduction is necessary when there are tens of thousands of candidate variables. To this end, \cite{fan08sis} introduced the Sure Independence Screening (SIS) method, which ranks the covariates based on the magnitude of their sample correlation with the response. More specifically, let $\bw=(w_1, w_2, \cdots, w_p)^T$ be a vector such that $w_j = |\hcorr(X_j, Y)|$ and $\gamma$ is a constant between $(0, 1)$, then a sub-model is defined as
\begin{equation}\label{eq:mgamma}
\calm_{\gamma} = \{j: w_j \text{ is amongst the largest }[\gamma n] \text{ of all} \},
\end{equation}  
where $[\gamma n]$ denotes the integer part of $\gamma n$. \cite{fan08sis} further demonstrated that SIS is screening consistent under some conditions. This guarantees that all those $X_j$'s with $\beta_j\ne0$ are included in the subset of covariates. 

To take advantage of the distribution information while implementing dimension reduction, we propose a new penalized variable selection approach that applies different penalties to each covariate based on the screening results. Let $\calm$ be the index set of covariates that have the largest $[n\backslash \log n]$ absolute sample correlation with the response among $X_1, X_2, \cdots, X_p$. We also define the set of paired covariates as
\begin{equation}
\calc = \{X_i: \exists j \text{ such that } (i,j)\in \calg \}.
\end{equation}

Our proposed method is established by solving the following optimization problem:
\begin{equation} \label{eq:penalty}
\min_{\bbeta\in\mathbb{R}^p} \frac{1}{2n}\|\by - X\bbeta\|_2^2
+\lambda_1\sum_{j:j\in \calc^c\cap\calm }|\beta_j| + \lambda_2 \sum_{j:j\in\calc\cap\calm}\beta_j^2
\end{equation}
subject to $\beta_j=0$ for $j\notin \calm$. In other words, we ignore the covariates that fail the marginal screening.

From the above penalty, it can be seen that we apply different penalties to covariates based upon the results from two types of screening. The intuitions behind the proposed penalty are  
\begin{itemize}
	\item
	For a covariate that is included in both $\calc$ and $\calm$, we only apply the $l_2$ penalty because it tends to be an important variable that we need to include in the final model.
	\item
	For a covariate that is included in $\calm$ but not in $\calc$, we only apply the $l_1$ penalty since there is no significant multicollinearity between it and other covariates. 
	\item 
	For a covariate that is not included in $\calm$, since it does not pass the marginal screening, we no longer consider it in the regression. This is because SIS enjoys screening consistency under certain assumptions, which implies that $\calm$ covers all important variables.
\end{itemize}

Our proposed method is connected with existing penalization approaches when the covariates have certain covariance structure. In particular, when the covariates are all independent, our method reduces to SIS-LASSO, which performs marginal screening first and then implements LASSO on the remaining covariates; when the predictors are all highly correlated such that $\calg$ includes all covariates pairs, our method is equivalent to SIS-Ridge.   

So far we have established a new penalized variable selection. Now we discuss how to solve the optimization problem in \eqref{eq:penalty}. One can see that the penalty part of \eqref{eq:penalty} is convex, so we can efficiently solve it by coordinate descent algorithm \citep{friedman10}. Specifically, the updating rule has the following form:
\begin{equation}
\hat{\beta}_j \leftarrow\left\{
\begin{aligned}
 & S(\frac{1}{N}\sum_{i=1}^Nx_{ij}(y_i-\tilde{y}_i^{(j)}), \lambda_1) &&\text{for }j\in\calc^c\cap\calm, \\ 
 & \frac{\frac{1}{N}\sum_{i=1}^Nx_{ij}(y_i-\tilde{y}_i^{(j)})}{1+\lambda_2} &&\text{for }j\in\calc\cap\calm,\\
\end{aligned}
\right.
\end{equation}
where $\tilde{y}_i^{(j)} = \hat{\beta}_0+\sum_{k\ne j}x_{ik}\hat{\beta}_k$ is the fitted value excluding the effect of $x_{ij}$, and $S(z) = \text{sign}(z)(|z|-\lambda)_{+}$ is the soft-thresholding function. In practice, we can first implement SIS to obtain $\calm$ when the dimension is high, then run the algorithm on the covariates $X_j$'s with $j\in\calm$.

\begin{remark}
	{The computational cost of the pairwise screening procedure is $O(p^2)$, which can be very inefficient as $p$ increases. In our proposed procedure, to reduce the computational complexity, we implement the marginal screening first to obtain $\calm$. Since the cardinality of $\calm$ is $O(n/\log(n))$, the computational cost of applying pairwise screening to $\calm$ will reduce to $O\big((n/\log(n))^2\big)$.}
\end{remark}

\subsection{Further extensions}\label{sec:extend}

As discussed in the previous subsection, we introduce a new penalized method that combines marginal screening with pairwise screening under the linear model setting.  Note that the pairwise covariates screening does not involve the response. Therefore, our method can be further extended to generalized linear models (GLM), e.g., logistic regression for binary response, or cox model for survival data. Suppose the response $Y$ is from the following one-parameter exponential family 
$f(y|\bx, \theta) = h(y)\exp\{y\theta-b(\theta)\}$. 
Moreover, we assume $\theta = \bx^T\bbeta$ for generalized linear models. 

Similar to \ref{eq:screen}, we define the pairwise screening as 
\begin{equation}
\calg_1 = \{(i,j): i < j, |\hcorr(X_i, X_j)|\ge a \}.
\end{equation}
The difference is that we do not consider the R squared screening for GLMs. This is because for GLMs, it is not reasonable to use the regression R squared to evaluate the associations between the covariates and the response. We further define the set of paired covariates as follows
\begin{equation}
\calc_1 = \{X_i: \exists j \text{ such that } (i,j)\in \calg_1 \}.
\end{equation}

Let $\displaystyle P_{\lambda_1, \lambda_2}(\bbeta)=\lambda_1\sum_{j:j\in \calc_1^c\cap\calm }|\beta_j| + \lambda_2 \sum_{j:j\in\calc_1\cap\calm}\beta_j^2$ be our proposed screening-based penalty. Then for logistic regression, we need to solve the following penalized maximum likelihood problem
\begin{equation}
\min_{\bbeta}\sum_{i=1}^n \big(y_i(\bx_i^T\bbeta)-\log(1+e^{\bx_i^T\bbeta}) \big)+P_{\lambda_1, \lambda_2}(\bbeta).
\end{equation}

In the above optimization problem, the log likelihood part can be approximated by a quadratic function, which is a weighted least squares term \citep{friedman10}. Therefore, it can still be solved by coordinate descent algorithm. Similarly, we can use the algorithm proposed by \cite{simon11cox} to solve the regularized Cox proportional hazard model using the screening based penalty $P_{\lambda_1, \lambda_2}(\bbeta)$. 

\section{Theoretical properties}\label{sec:theo}

In this section, we study the theoretical properties of the proposed pairwise correlation screening (PCS) method. More specifically, we investigate the conditions under which the PCS achieves the variable selection consistency. 

Note that we implemented the marginal screening using SIS to the covariates set. Fan and Lv \cite{fan08sis} demonstrated that under certain regularity conditions, SIS has the screening consistency, that is, the resulting subset of covariates includes all important variables. {Due to space constraints, we only present the main result. The regularity conditions $(A1)-(A4)$ are provided in the appendix.}

\begin{proposition}[\cite{fan08sis}]\label{thm:sis}
	Under $(A1)-(A4)$, if $2\kappa + \tau <1$, then there is some  $\theta < 1-2\kappa -\tau$ such that , when $\gamma\sim cn^{-\theta}$ with $c>0$, we have, for some $C>0$,
	\begin{equation}
	\PP{\calm^* \subset \calm_{\gamma}} =  1-O[\exp\{-C^{1-2\kappa}/\log(n)\}],
	\end{equation}
	where $\calm_{\gamma}$ is the subset of covariates obtained from the sure independence screening. 
\end{proposition}

The above proposition guarantees that all important variables survive the marginal screening with high probability. In order to achieve the selection consistency, we also need to ensure that only important variables can pass the pairwise screening. In the following theorem, we present the technical conditions that are required such that the event $\calc\cap\calm \subset \calm^*$ occurs with high probability.

\begin{theorem}\label{thm:include}
	Suppose the following conditions holds
	\begin{enumerate}[label=(\subscript{B}{{\arabic*}})]
		\item
		$ n/p^2 \rightarrow 0$. 
		\item
		There exists $\eta >0$ such that either one of the following two conditions holds:
		\begin{enumerate}
			\item
			$\limn \log p/n\rightarrow \eta_0, \max_{i\in\calm^*, j\in\calm\backslash\calm^*}|\tcorr(X_i, X_j)| < \min\{\eta, 1-e^{-4\eta_0}\} $
			\item
			$\limn \log p/n \rightarrow 0, \max_{i\in\calm^*, j\in\calm\backslash\calm^*}|\tcorr(X_i, X_j)| < \eta$. 
		\end{enumerate}	
	\end{enumerate}
	Here $\tcorr(X_i, X_j)$ denotes the population correlation between covariates $X_i$ and $X_j$. Then under conditions $(B1)$ and $(B2) (a)$ or conditions $(B1)$ and $(B2) (b)$, we have that as $\infn$,
	\begin{equation}
	\PP{\calc\cap\calm \subset \calm^*}\rightarrow 1. 
	\end{equation}
\end{theorem}

{Given Proposition \ref{thm:sis} and Theorem \ref{thm:include}, to demonstrate the selection consistency of PCS, we only need to show that the $l_1$ penalty in \eqref{eq:penalty} can identify the important variables in $\calc^c\cap\calm$ exactly. This relates to the selection consistency for the LASSO, which has been studied extensively. In particular, Zhao and Yu \cite{zhao06} have shown that the Irrepresentable Condition (to be clarified later) is almost necessary and sufficient for LASSO to select all important variables. }

{We first introduce some necessary notations.} Let $C = \frac{1}{n}X^TX$. Without of loss of generality, assume that $\bbeta = (\beta_1, \beta_2, \ldots, \beta_p)^T$ where $\beta_j\ne 0$ for $j=1, \ldots, s$ and $\beta_j=0$ otherwise. By Theorem \ref{thm:include}, we further assume that $\calc\cap \calm =\{1, \ldots, s_1 \} $ where $1\le s_1\le s$. Then the design matrix $X$ can be expressed as $X = (X^{1}_{(1)}, X^{2}_{(1)}, X_{(2)})$, where $X^{1}_{(1)}$ corresponds to the first $s_1$ columns, $X^{2}_{(1)}$ corresponds to the $(s_1+1)$th to the $s$th columns and  $X_{(2)}$  corresponds to the last $p-s$ columns of $X$ respectively. Similarly, we can write $\bbeta_1^{(1)} = (\beta_1, \ldots, \beta_{s_1})^T$, $\bbeta^{(1)}_2 = (\beta_{s_1+1}, \ldots, \beta_s)^T$, and $\bbeta^{(2)} = (\beta_{s+1}, \ldots, \beta_{p})^T$.

Set $C_{11}^{(11)} = \frac{1}{n}{X^{1}_{(1)}}^TX^{1}_{(1)}$, $C_{11}^{(12)} =\frac{1}{n} {X^{1}_{(1)}}^TX^{2}_{(1)}$, $C_{11}^{(21)} = \frac{1}{n}{X^{2}_{(1)}}^T X^{1}_{(1)}$, $C_{11}^{(22)} = \frac{1}{n} {X^{2}_{(1)}}^TX^{2}_{(1)}$, $C^{(1)}_{21} = \frac{1}{n} X_{(2)}^T X^{1}_{(1)} $ , $C^{(2)}_{21} = \frac{1}{n}X_{(2)}^T X^{2}_{(1)}  $, $C_{22} = \frac{1}{n}X_{(2)}^TX_{(2)}$, $C^{(1)}_{12} = \frac{1}{n}  {X^{1}_{(1)}}^T X_{(2)}$, $C^{(2)}_{12} = \frac{1}{n}  {X^{2}_{(1)}}^T X_{(2)}$. Then $C$ can be expressed in a block-wise form as follows:
$$
\begin{pmatrix}
C_{11}^{(11)} & C_{11}^{(12)} & C^{(1)}_{12} \\
C_{11}^{(21)} & C_{11}^{(22)} & C^{(2)}_{12}  \\
C^{(1)}_{21}  & C^{(2)}_{21}   & C_{22}
\end{pmatrix}
$$

We impose the following assumption analogous to the  Irrepresentable Condition introduced by Zhao and Yu \cite{zhao06}. Specifically, we assume that there exists a constant $\delta>0$, such that 
\begin{equation}\label{eq:irr}
\| C^{(2)}_{21} {(C_{11}^{(22)}})^{-1}\text{sign}(\bbeta_2^{(1)})\|_{\max}\le 1-\delta,
\end{equation}
where $\| \cdot \|_{\max}$ is the max norm.

In fact, we can show that the condition mentioned above is implied by the Irrepresentable Condition on the full covariates set $\calm$ under mild assumptions. We illustrate this result in the following theorem:

\begin{theorem}\label{thm:condition}
	Assume that there exists $\lambda_0>0$ so that $\lambda_{min}(C_{11}^{(11)})\ge \lambda_0$, $\lambda_{min}(C_{11}^{(22)})\ge \lambda_0$, and conditions $(B1)$ and $(B2) (b)$ holds. Suppose the Irrepresentable Condition holds, i.e., $\exists \xi >0$ s.t.
	\begin{equation}\label{eq:zy}
	\|C_{21}C_{11}^{-1}\text{sign}(\bbeta_1)\|_{\max}\le 1-\xi,
	\end{equation}
  where $C_{11} = \begin{pmatrix}
	C_{11}^{(11)} & C_{11}^{(12)} \\
	C_{11}^{(21)} & C_{11}^{(22)} \\
	\end{pmatrix}$, $C_{21} = \begin{pmatrix}
	C_{21}^{(1)} & C_{21}^{(2)}
	 \end{pmatrix}$ , $\bbeta_1 = (\beta_1, \ldots, \beta_s)^T$  and $\xi$ is a positive constant. $\lambda_{min}(\cdot)$ Then with probability tending to 1, the condition \eqref{eq:irr} holds.
\end{theorem}
 
{The assumptions $\lambda_{min}(C_{11}^{(11)})\ge \lambda_0$, $\lambda_{min}(C_{11}^{(22)})\ge \lambda_0$ in Theorem 4 require that $C_{11}^{(11)}$ and $C_{11}^{(22)}$ have eigenvalues bounded below. 
Given the Irrepresentable Condition in \eqref{eq:zy}, we need additional constraints on the random noise $\veps_i$'s and the coefficients of important variables $\beta_1,\cdots, \beta_s$.}

\begin{enumerate}[label=(\subscript{C}{{\arabic*}})]
	\item 
	{$\veps_i$'s are i.i.d. random variables with finite $2k$'s moment $E(\veps_i)^{2k}<\infty$ for an integer $k>0$.}
	\item
	{There exists $0<\alpha \le 1$ and $d_0>0$ such that $n^{\frac{1-\alpha}{2}}\min_{j=1, \cdots, s}|\beta_j|\ge d_0$.}
\end{enumerate}

So far we have discussed all the theoretical assumptions required to ensure the selection consistency of the proposed PCS method. We conclude the consistency result in the following theorem:

\begin{theorem}
	Suppose conditions (A1)--(A4), (C1)--(C2) and inequality \eqref{eq:zy} hold, and the assumptions of Theorem \ref{thm:condition} are satisfied, then for any $\lambda_1$ such that $\frac{\lambda_1}{\sqrt{n}}=o(n^{\alpha}/2)$ and $\frac{1}{p}(\frac{\lambda_1}{\sqrt{n}})\rightarrow\infty$, we have,
	\begin{equation}
	\PP{\{j:\hat{\beta}_j\ne 0\} = \calm^* }\rightarrow 1 \text{ as } n\rightarrow\infty,
	\end{equation}
where $\hat{\bbeta} = (\hat{\beta_1}, \ldots, \hat{\beta_p})^T$ is the solution to \eqref{eq:penalty}.	
\end{theorem}

The proof follows immediately from Proposition 2 and Theorems 3 and 4 {as well as the selection consistency of the LASSO}. It shows that under certain conditions, our proposed method is consistent in variable selection. In Section \ref{sec:pairns}, we will show with numerical examples that our proposed method can perform well in practice.

\section{Numerical Studies}\label{sec:pairns}

In Section 3, we have established a new regularized variable selection approach for high-dimensional linear models. In this section, we demonstrate the performance of our proposed method using both simulations and real data examples.  

\subsection{Simulation study}\label{sec:sim}

In this section, we use several simulated examples to show that our method with pairwise correlation screening (PCS) or pairwise rank-based correlation screening (PRCS) outperforms some existing variable selection procedures. More specifically, PCS denotes our proposed method using the limiting distribution in Theorem \ref{thm:wdist}, and PRCS uses the asymptotic result in Proposition \ref{thm:spear}. 

For comparison, we consider LASSO, elastic net (Enet), SIS-LASSO, SIS-elastic net (SIS-Enet) {and SIS-PACS. The SIS-PACS refers to applying the PACS method proposed by \cite{sharma13} after implementing the SIS procedure}. In SIS-type methods, we first implement SIS and find those covariates with the largest $[n\backslash \log n]$ absolute sample correlations with the response, then perform LASSO, Enet or PACS on these variables. We evaluate the variable selection accuracy using False Negatives (FN) and False Positives (FP). 
FN is defined as
$
FN = \sum_{j=1}^p I(\hbeta_j=0)\times I(\beta_j\ne0),  
$
where $I(\cdot)$ denotes the indicator function, and FP is defined as 
$
FP = \sum_{j=1}^p I(\hbeta_j\ne0)\times I(\beta_j=0).  
$
We use the following quantities to evaluate the prediction accuracy:

\begin{itemize}
	\item 
	$\|\hat{\bbeta}-\bbeta_0\|_2$: the $l_2$ distance between the estimated coefficient vector and the true coefficients $\bbeta_0$;
	\item
	Out of sample mean squared errors (MSE) on the independent test data;
\end{itemize}

\begin{table}[!t]
	\centering
	\def\arraystretch{1.02}
	\caption{Results for simulated Example 1. For each method, we report the average MSE, $l_2$ distance, FN and FP over 100 replications (with standard errors given in parentheses).}
	\begin{tabular}{lcccr}\hline
		\noalign{\vskip 3pt}
		Method             &  MSE & $\|\hat{\bbeta}-\bbeta_0\|_2$  & FN & \multicolumn{1}{c}{FP} \\
		\multicolumn{5}{c}{$p=1000, \quad \sigma = 2$} \\
		Elnet & 5.94 (0.07) & 1.40 (0.03) & 0.00 (0.00) & 1.64 (0.24)\\
		SIS-Elnet & 5.47 (0.06) & 1.30 (0.03) & 0.00 (0.00) & 1.15 (0.12) \\
		LASSO & 5.95 (0.07) & 1.50 (0.03) & 0.00 (0.00) & 1.28 (0.18) \\
		SIS-LASSO & 5.47 (0.06) & 1.42 (0.03) & 0.00 (0.00) & 0.85 (0.10) \\
		SIS-Ridge & 86.00 (0.76) & 4.50 (0.01) & 0.00 (0.00) & 12.00 (0.00) \\
	    SIS-PACS & 4.69 (0.07) &  0.48 (0.02) & 0.00 (0.00) & 0.01 (0.01) \\		
		PCS & 4.74 (0.05) & 0.76 (0.02) & 0.00 (0.00) & 0.03 (0.02) \\
		PRCS & 4.91 (0.05) & 0.93 (0.02) & 0.00 (0.00) & 2.55 (0.15) \\
		\multicolumn{5}{c}{$p=5000, \quad \sigma = 2$} \\
		Elnet & 6.42 (0.09) & 1.57 (0.03) & 0.00 (0.00) & 2.45 (0.26) \\
		SIS-Elnet & 5.64 (0.06) & 1.41 (0.03) & 0.00 (0.00) & 1.28 (0.12) \\
		LASSO & 6.41 (0.08) & 1.64 (0.04) & 0.00 (0.00) & 2.06 (0.21) \\
		SIS-LASSO & 5.65 (0.06) & 1.52 (0.03) & 0.00 (0.00) & 1.03 (0.10) \\
		SIS-Ridge & 88.74 (0.75) & 4.59 (0.01) & 0.00 (0.00) & 12.00 (0.00) \\
		SIS-PACS & 4.97 (0.08) & 0.72 (0.02) & 0.00 (0.00) & 1.78 (0.43) \\
		PCS & 4.77 (0.05) & 0.81 (0.03) & 0.00 (0.00) & 0.02 (0.02) \\
		PRCS & 4.85 (0.06) & 0.89 (0.03) & 0.00 (0.00) & 1.21 (0.11)\\	
		\hline
	\end{tabular}\label{tab:sim1}
\end{table}

\begin{table}[!t]
	\centering
	\def\arraystretch{1.02}
	\caption{Results for simulated Example 2. The format of this table is the same as Table \ref{tab:sim1}.}
	\begin{tabular}{lcccr}\hline
		\noalign{\vskip 3pt}		
		Method             &  MSE & $\|\hat{\bbeta}-\bbeta_0\|_2$  & FN & \multicolumn{1}{c}{FP} \\
		\multicolumn{5}{c}{$p=1000, \quad \sigma = 2$} \\		
		Enet         & 6.75 (0.08) & 2.45 (0.02) & 1.00 (0.01) & 0.98 (0.25) \\
		SIS-Enet & 6.47 (0.10) & 2.30 (0.03) & 0.76 (0.05) & 3.16 (0.41)\\
		LASSO & 6.75 (0.08) & 2.45 (0.02) & 1.00 (0.01) & 0.98 (0.25) \\
		SIS-LASSO & 6.47 (0.10) & 2.30 (0.03) & 0.76 (0.05) & 3.16 (0.41) \\
		SIS-Ridge & 14.14 (0.10) & 3.85 (0.00) & 0.27 (0.04) & 19.27 (0.04) \\
		SIS-PACS & 6.53 (0.14) & 2.43 (0.04) & 1.06 (0.05) & 3.39 (0.73) \\
		PCS & 5.24 (0.12) & 1.41 (0.08) & 0.34 (0.05) & 1.63 (0.13)\\	
		PRCS & 5.72 (0.13) & 1.75 (0.08) & 0.43 (0.05) & 1.34 (0.24)\\	
		\multicolumn{5}{c}{$p=5000, \quad \sigma = 2$} \\	
		Elnet & 7.16 (0.08) & 2.55 (0.02) & 1.02 (0.01) & 0.40 (0.09) \\
		SIS-Elnet & 7.02 (0.09) & 2.49 (0.03) & 0.94 (0.03) & 1.31 (0.34) \\
		LASSO & 7.16 (0.08) & 2.55 (0.02) & 1.02 (0.01) & 0.36 (0.08) \\
		SIS-LASSO & 7.03 (0.09) & 2.49 (0.03) & 0.94 (0.03) & 1.31 (0.34) \\
		SIS-Ridge & 14.40 (0.11) & 3.87 (0.00) & 0.59 (0.05) & 19.59 (0.05) \\
		SIS-PACS & 7.28 (0.16) & 2.83 (0.04) & 1.26 (0.07) & 2.41 (0.95) \\
		PCS & 5.96 (0.14) & 1.83 (0.09) & 0.63 (0.06) & 0.74 (0.08) \\	
		PRCS & 6.48 (0.13) & 2.14 (0.07) & 0.68 (0.05) & 0.73 (0.24) \\							
		\hline
	\end{tabular}\label{tab:sim2}
\end{table}

\begin{table}[!t]
	\centering
	\def\arraystretch{1.15}
	\caption{Results for simulated Example 3. The format of this table is the same as Table \ref{tab:sim1}.}
	\begin{tabular}{lcccr}\hline
		\noalign{\vskip 3pt}
		Method             &  MSE  & $\|\hat{\bbeta}-\bbeta_0\|_2$  & FN & \multicolumn{1}{c}{FP} \\
		
		Enet         & 69.71 (0.88) & 5.13 (0.03) & 4.99 (0.13) & 1.57 (0.37) \\
		SIS-Enet & 72.54 (0.88) & 5.25 (0.03) & 5.65 (0.10) & 0.23 (0.12) \\
		LASSO & 72.78 (0.87) & 5.41 (0.03) & 6.06 (0.10) & 0.09 (0.04) \\
		SIS-LASSO & 70.12 (0.86) & 5.35 (0.04) & 5.69 (0.12) & 0.94 (0.19) \\
		SIS-Ridge & 109.66 (0.87) & 5.74 (0.01) & 4.46 (0.06) & 16.46 (0.06) \\
		SIS-PACS & 71.27 (0.89) & 5.58 (0.02) & 5.06 (0.02) & 3.45 (0.07) \\
		PCS & 58.87 (0.50) & 4.80 (0.04) & 4.95 (0.03) & 0.06 (0.06) \\
		PRCS & 59.76 (0.56) & 4.83 (0.04) & 4.97 (0.02) & 0.00 (0.00)	 	\\	
		\hline
	\end{tabular}\label{tab:sim3}
\end{table}

\begin{table}[!t]
	\centering
	\def\arraystretch{1.15}
	\caption{Results for simulated Example 4. The format of this table is the same as Table \ref{tab:sim1}.}
	\begin{tabular}{lcccr}\hline
		\noalign{\vskip 3pt}		
		Method             &  Classification Error & $\|\hat{\bbeta}-\bbeta_0\|_2$  & FN & \multicolumn{1}{c}{FP} \\
		Enet         & 0.129 (0.003) & 5.79 (0.01) & 2.16 (0.17) & 12.77 (1.54) \\
		SIS-Enet & 0.126 (0.003) & 5.69 (0.03) & 1.37 (0.15) & 7.48 (0.39)\\
		LASSO & 0.136 (0.003) & 5.83 (0.01) & 4.19 (0.13) & 4.25 (0.49) \\
		SIS-LASSO & 0.130 (0.003) & 5.75 (0.02) & 3.94 (0.12) & 3.50 (0.32) \\
		SIS-Ridge & 0.311 (0.003) & 6.28 (0.01) & 0.11 (0.05) & 12.11 (0.05) \\
		PCS & 0.098 (0.004) & 5.39 (0.05) & 1.73 (0.14) & 2.92 (0.31)\\	
		PRCS & 0.099 (0.004) & 5.34 (0.06) & 1.71 (0.13) & 3.26 (0.32) \\		
		\hline
	\end{tabular}\label{tab:sim4}
\end{table}

\begin{table}[!t]
	\centering
	\def\arraystretch{1.15}
	\caption{Results for simulated Example 5. The format of this table is the same as Table \ref{tab:sim1}.}
	\begin{tabular}{lcccr}\hline
		\noalign{\vskip 3pt}
		Method             &  MSE  & $\|\hat{\bbeta}-\bbeta_0\|_2$  & FN & \multicolumn{1}{c}{FP} \\
		Enet         & 102.47 (1.84) & 3.90 (0.08) & 1.51 (0.12) & 4.88 (0.86) \\
		SIS-Enet & 96.60 (2.74) & 3.49 (0.09) & 1.02 (0.12) & 4.20 (0.37) \\
		LASSO & 103.11 (1.89) & 4.42 (0.08) & 2.30 (0.13) & 3.74 (0.71) \\
		SIS-LASSO & 96.97 (2.78) & 4.27 (0.08) & 2.05 (0.14) & 1.87 (0.20)\\
		SIS-Ridge & 226.52 (3.78) & 4.95 (0.03) & 0.26 (0.08) & 12.26 (0.08) \\
		SIS-PACS & 89.82 (2.70) & 3.54 (0.15) & 0.26 (0.08) & 7.32 (0.45) \\
		PCS & 79.79 (3.16) & 2.42 (0.14) & 0.42 (0.10) & 1.29 (0.33) \\
		PRCS & 74.60 (1.24) & 2.15 (0.12) & 0.31 (0.08) & 0.06 (0.03)	 	\\	
		\hline
	\end{tabular}\label{t:sim5}
\end{table}

We generate the simulated data from Model \eqref{eq:pairlm} and conduct 100 replications. Each simulated dataset includes a training set of size 100, an independent validation set of size 100 and an independent test set of size 400. {Here we fix the sample size to be 100 throughout the simulation study. In the next subsection, we also consider varying sample size for sensitivity study.} We only fit models on the training data, and we use the validation data to select tuning parameters. {Given the fitted model, we can calculate the FN, FP and the estimation error $\|\hat{\bbeta}-\bbeta_0\|_2$, and we make predictions and calculate the out of sample MSEs using the test data.} We simulate the covariates from the multivariate Gaussian distribution $\caln(0, \Sigma)$, with $\Sigma = (\sigma_{ij})_{p\times p}$ being the correlation matrix.

Details of the simulated examples are as follows:

{\bf Example 1:} {We consider $p=1000$ or $5000$, $\sigma =2$}, and we take $\bbeta = (2, 2, \cdots, 2, 0, \cdots, 0)^T$ where the first 10 coefficients being non-zero and equal to 2. We set $\sigma_{ij}=0.8$ for $1\le i\ne j\le 5$, $6\le i\ne j\le 10$ and $0$ for all the other $i\ne j$. We also consider $\sigma = 6$ and present the results in the supplementary. In other words, there are two groups in the covariates, where each group has 5 important variables. 

{\bf Example 2:} {We consider $p=1000$ or $5000$, $\sigma =2$}, $\bbeta_0=(3,-1.5, 2, 0, \cdots, 0, \cdots, 0)^T$, where the first 3 coefficients are non-zero ones. We also consider $\sigma = 6$ and present the results in the supplementary. We generated Gaussian covariates with $\sigma_{ij} = 0.5^{|i-j|}$ for $1\le i\ne j\le 1000$.  

{\bf Example 3:} The coefficients have the same set up as in Example 1. But we set $\sigma_{ij}=0.8$ for $1\le i\ne j\le 5$ and $0$ for all the other $i\ne j$. Therefore only part of the important variables are highly correlated. {We consider $p=5000$ and $\sigma = 6$ in this Example.}

{\bf Example 4:} In this example, we examine the performance of all methods under the logistic regression setting. We simulate the binary response $Y$ from the binomial distribution $\text{Binom}(1, \frac{\exp\{X^T\bbeta+\sigma\}}{1+\exp\{X^T\bbeta+\sigma\}})$, where $X$, and $\bbeta$ follow the same set ups as in Example 1. {We consider $p=5000$ and $\sigma = 6$ in this Example.} Instead of comparing MSE, we calculate the classification errors on the test data. {We did not compare with SIS-PACS in this example since the R program does not support GLM.}

{\bf Example 5:} In this example, we generate the covariates from a multivariate $t$ distribution, where $X_j$'s are $t$ distributed with degrees of freedom 5. The covariance structure of the covariates and the coefficients are set the same as in Example 1. {We consider $p=5000$ and $\sigma = 6$ in this Example.}

The results for simulated Example 1 is shown in Table \ref{tab:sim1}. We see that when there are groups in the covariates, the performance improvement of our approach is significant compared with other penalized methods. While elastic net-based procedures perform better than LASSO-type approaches in terms of FN, as illustrated by \cite{zou05elnet}, they still miss approximately one important covariate on average. In contrast, the model selection results of our method are much closer to the correct model for this example. {In addition, although SIS-PACS has competitive performance when $\sigma$ is small, it tends to include more unimportant variables into the model when the noise level increases, and therefore may not work well.}

Table \ref{tab:sim2} displays the performance comparisons for Example 2. Compared with Example 1, this setting is a more difficult one for our method, since correlation exists among all pairs of covariates. Nevertheless, PCS and PRCS perform better than, or as well as all the others in terms of estimation error and prediction accuracy. Moreover, besides SIS-Ridge, our proposed methods are able to identify more important variables than others in this example {when the noise level is low}. 

Table \ref{tab:sim3} shows the results for Example 3, where correlation exists only within part of the important variables. This example is more difficult compared with Example 1 due to the correlation structure of the covariates. One can see that the false negatives are significantly larger for all procedures. Nevertheless our method still outperforms all the others in terms of prediction and variable selection accuracy. 

Example 4 considers the logistic regression setting, and the results are provided in Table \ref{tab:sim4}. One can see that as the correlations among the covariates vary, the performance of our method is always competitive compared with the others. 

Table \ref{t:sim5} displays the results for all methods under the non-Gaussian covariates setting. Similar to Example 1, our proposed PCS and PRCS achieve much better performance compared with the competitors. Moreover, due to the non-Gaussian set ups, the nonparametric method PRCS outperforms PCS.

As a conclusion, our method can make use of the correlation structure among predictors. Compared with other penalized variable selection procedures, our method performs well, especially when the covariates are highly correlated. 

\subsection{Sensitivity Study}

\begin{figure*}[!t]
	\centering
	\subfigure[ MSE]{\includegraphics[width=.4\textwidth]{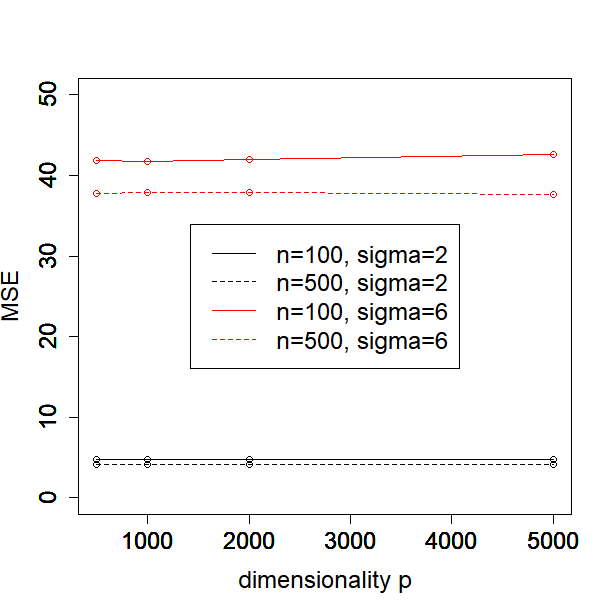}}
	\subfigure[Estimation error]{\includegraphics[width=.4\textwidth]{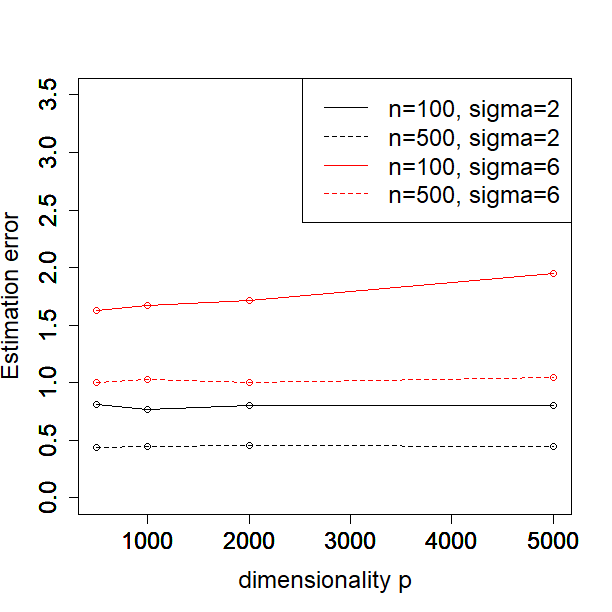}}
		\subfigure[ FN]{\includegraphics[width=.4\textwidth]{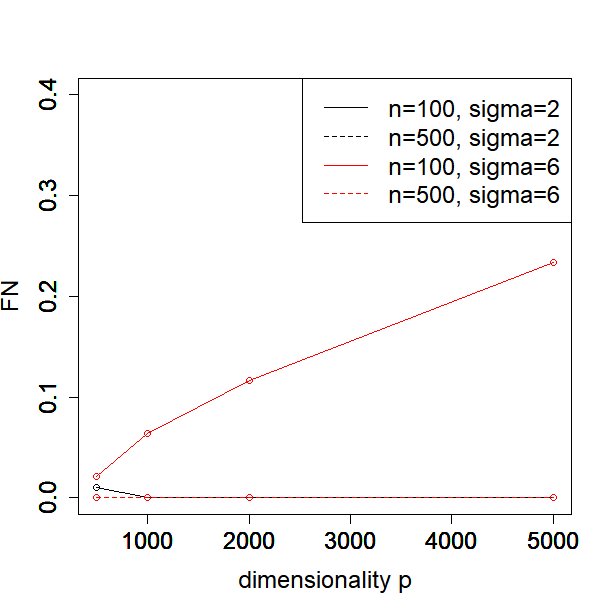}}
			\subfigure[ FP]{\includegraphics[width=.4\textwidth]{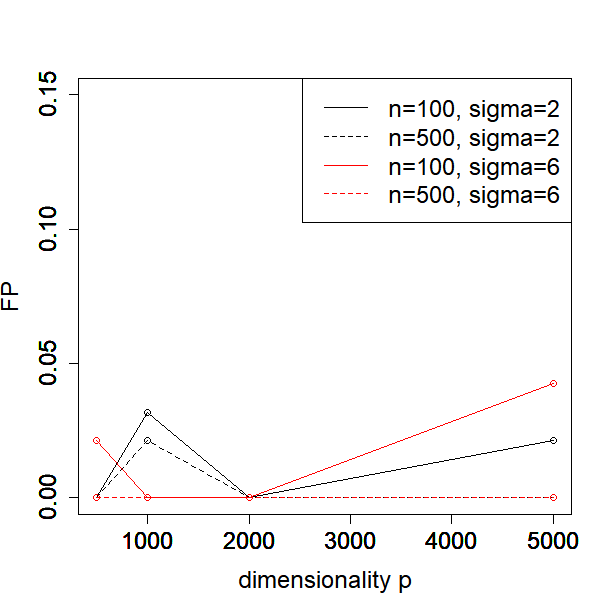}}
	\caption{Performance of PCS against different dimensionality $p$.} 
	\label{fig:sens}
\end{figure*}

In this subsection, we investigate how the performance of our method depends on the sample size, dimensionality, and noise level. In particular, we consider $n=100$ or $500$, $p=500, 1000, 2000$ or $5000$ and $\sigma  = 2$ or $6$ in the Simulated Example 1 as introduced in Section~\ref{sec:sim}. We illustrate the MSE, $\|\hat{\bbeta}-\bbeta_0\|_2$,  FN and FP against different values of $p$ for each configuration of sample size and noise level in Figure \ref{fig:sens}.

One can see from the plots that the performance of PCS does not change much as the dimensionality $p$ increases from 500 to 5000, especially in terms of MSE and the estimation error of $\bbeta_0$. Moreover, the performance is better when the sample size and signal to noise ratio (SNR) become larger, which is expected. In general, our proposed PCS method is robust to sample size, dimensionality and SNR.

\subsection{Soil data}

\begin{figure}[!t]
	\centering
	\includegraphics[width = .6\textwidth]{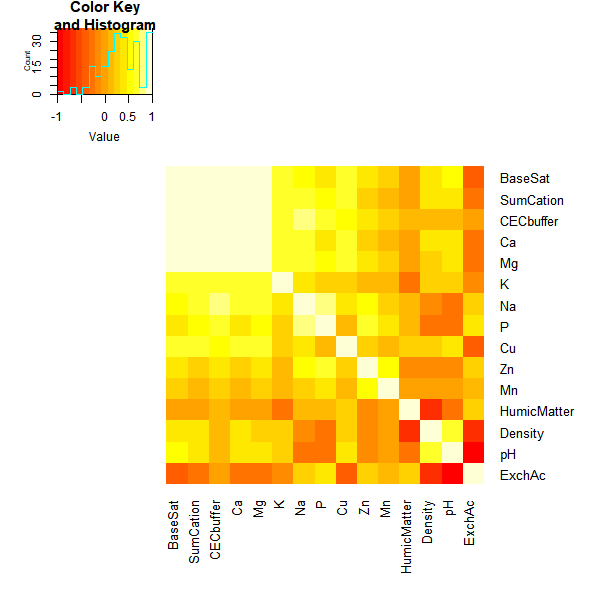}
	\caption{Heatmap for the absolute pairwise correlation matrix of the covariates for the soil data.}\label{fig:soil}
\end{figure}

We first demonstrate the performance of our method in real applications using a small dataset. This dataset contains 15 covariates of soil characteristics for 20 plots with the same area in the Appalachian Mountain. The outcome variable is the forest diversity for each plot. More descriptions of the data can be found in \cite{bondell08}. To better demonstrate the correlation structure of covariates, we obtain the absolute pairwise correlation matrix and show the heatmap in Figure \ref{fig:soil}. One can see that some predictors are highly correlated. In particular, the magnitude of the pairwise correlations among Sum of Cations (SumCation), calcium, magnesium, Base Saturation (BaseSat), and cation exchange capacity (CEC) are as large as 0.9. The reason is SumCation, BaseSat, CEC are characteristics for cations; while calcium and magnesium are examples of cations \citep{bondell08}. 

\begin{table}[!b]
	\centering
	\begin{tabular}{lcr}	\hline
		\noalign{\vskip 3pt}
		Method             &  MSE  & \multicolumn{1}{c}{ Model Size }\\
		
		Enet & 1.088 (0.047)   & 3.70 (0.38)  \\
		LASSO & 1.068 (0.045) & 2.08 (0.21) \\
		Ridge & 1.113 (0.044)  & 15.00 (0.00)     \\
		PCS & 0.996 (0.062) 	& 5.82 (0.37) 	 \\	
		PRCS & 1.028 (0.063)   &  5.96 (0.38) \\
		\hline
	\end{tabular}
	\caption{Average mean squared errors and model size (with standard errors in parenthesis) for Enet, LASSO, Ridge and our methods on the soil data.}\label{tab:soil}
\end{table}

We conduct a total of 100 replications. In each replication, 15 samples are randomly chosen as the training set and the remaining as the test set. As in the simulation experiments, we applied LASSO, Enet, Ridge and our proposed PCS, PRCS to the dataset. For each method, 5-fold cross-validation is used to choose the tuning parameters {since the sample size is very small}. We report the average prediction errors on the test data and the model size in Table \ref{tab:soil}. One can see that PCS and PRCS outperform all the others in terms of prediction accuracy. Moreover, PCS and PRCS tend to include more covariates into the model compared with LASSO and Enet. 

To further investigate the performance of variable selection, we summarize the frequency that each covariate is selected for LASSO, Enet and our method, which is displayed in Table \ref{tab:freq}. Note that among those variables that are most frequently selected by LASSO and Enet, for instance, CEC, Mn, HumicMatt, they also tend to be included for our method. Moreover, our method can identify covariates that are strongly correlated. For example, potassium, sodium and copper are variables related to cations, and all have a large sample correlation with CEC, which is a potentially important variable. These variables are frequently selected by our method, but not by Enet and LASSO. 

\begin{table}[!t]
	\centering
	\begin{tabular}{lccc}
		& PCS & Enet & LASSO  \\
	Variables & & &   \\\hline
	BaseSat & 16 & 9 &  0\\
	SumCaton & 32 & 23 & 0 \\
	CECbuffer & 86 & 62 & 48  \\
	Ca & 37 & 32 & 11  \\
	Mg & 6 & 10 & 0  \\
	K & 49 & 27 & 12 \\
	Na & 22 & 10 & 6 \\
	P & 32 & 15 &  5 \\	
	Cu & 47 & 17 & 9 \\
	Zn & 29 & 17 & 4 \\
	Mn & 69 &  43 &  32 \\	
	HumicMatt & 89 & 70 & 69 \\	
	Density & 25 & 15 & 4 \\
	pH & 27 &  11 & 4 \\		
	ExchAc & 16 & 9 & 4  \\
    \hline
	\end{tabular}
	\caption{Frequency of each variable being selected for PCS, Enet and LASSO out of 100 replications.}\label{tab:freq}
\end{table}

\subsection{Riboflavin data}

In this section, we consider a real data set about the riboflavin production in Bacillus subtilis. The data contain $n=71$ samples, where the response variable is the logarithm of the riboflavin production rate, and the covariates are the logarithm of expression levels of $p=4081$ genes. More descriptions about the dataset can be found in \cite{buhlmann14high}. Before analysis, all covariates are standardized to have zero means and unit standard deviations. 

For comparison purpose, we apply LASSO, Enet, SIS-LASSO, SIS-Enet, SIS-ridge and our method to the dataset. We conduct 100 replications, and we randomly split the dataset into a training set of size $50$ with the remaining as the test data. For all methods, we implement 10-fold cross validation on the training data to select the penalty parameters. 

The results are reported in Table \ref{tab:ribof}. One can see that PCS has significant improvement in terms of out of sample mean squared errors compared with other competitors. On the other hand, PRCS does not perform well compared with PCS. A possible reason is that in this dataset all the variables have been taken log transformations and are approximated well by Gaussian distribution. Moreover, due to the assumption of Proposition \ref{thm:spear} where $\log p =o(n^{1/3})$, PRCS is more sensitive to the dimensionality and the sample size of dataset. As a result, PRCS may not achieve good performance when the dimensionality is too high.  

We also examine the gene selection results. There are 8 genes that are selected at least 50 times out of the 100 replications by our method, i.e., \text{XTRA\_at}, {YCKE\_at},  {YDAR\_at},  {YOAB\_at}, {YWFO\_at}, {YXLC\_at}, {YXLD\_at} and {YXLE\_at}. Besides {YXLC\_at}, all the other genes also appear among the most frequently selected genes by SIS-Enet and SIS-LASSO with a frequency no less than 50. For  {YXLC\_at}, we find that the magnitude of the pairwise sample correlations between this gene and two other genes, {YXLD\_at} and {YXLE\_at}, are greater than 0.95. It indicates that our method is capable of identifying potentially important variables that are highly correlated with the others. 

\begin{table}[!t]
	\centering
	\begin{tabular}{lcr}\hline
		\noalign{\vskip 3pt}		
		Method             &  MSE  & Model Size \\		
		SIS-Enet & 0.358 (0.015) & 15.66 (0.46)  \\
		SIS-Lasso & 0.356 (0.016) & 9.12 (0.18) \\
		SIS-Ridge & 0.632 (0.024)  & 26.00 (0.00)     \\
		PCS & 0.327 (0.014) 	& 15.04 (0.39) 	 \\
		PRCS & 0.361 (0.018)	& 12.77 (0.37) \\
		\hline
	\end{tabular}
	\caption{Average mean squared errors and model size (with standard errors in parenthesis) for SIS-Enet, SIS-LASSO, SIS-Ridge, PCS and PRCS applied to the riboflavin data.}\label{tab:ribof}
\end{table}

\section{Discussion}
\label{sec:pairsum}

In summary, we propose a novel variable selection method that regularizes covariates selectively based on the results from two screening procedures: pairwise screening and marginal screening. The screening process of covariates pairs takes advantage of the distribution information of the maximal absolute pairwise sample correlation among covariates, and is applicable to large scale problems. Simulation experiments and real data study demonstrate that the proposed method performs well when important variables are highly correlated compared with existing approaches. For future research, we can consider other extensions of our proposed method, for example, the Cox model for survival data. 

\appendix
\section{Technical Proofs}
\input{appendix.tex}

\small
\bibliography{vsbib}
\bibliographystyle{plainnat}

\vskip .45cm
\noindent
Siliang Gong
\vskip 2pt
\noindent
University of Pennsylvania
\vskip 2pt
\noindent
E-mail: (siliang@pennmedicine.upenn.edu)
\vskip 2pt

\noindent
Kai Zhang
\vskip 2pt
\noindent
The University of North Carolina at Chapel Hill
\vskip 2pt
\noindent
E-mail: (zhangk@email.unc.edu)
\vskip 2pt

\noindent
Yufeng Liu
\vskip 2pt
\noindent
The University of North Carolina at Chapel Hill
\vskip 2pt
\noindent
E-mail: (yfliu@email.unc.edu)

\end{document}

%% file: appendix.tex

\hspace{12pt}
We present some regularity conditions and key proofs in the appendix. 

\noindent{\bf Regularity Conditions for Sure Independence Screening}
	Define $\bz = \Sigma^{-1/2}\bx$, $Z = X\Sigma^{-1/2}$. Let $\calm^*$ be the index set of covariates with non-zero coefficient.  The following assumptions are imposed:
	\begin{enumerate}[label=(\subscript{A}{{\arabic*}})]
		\item 
		$p>n$ and $\log(p) = O(n^{\epsilon})$ for some $\epsilon\in(0, 1-2\kappa)$, where $\kappa$ is given by condition (A3).
		\item
		$\bz$ has a spherically symmetric distribution, and  $\exists c_0, c_1>1, C_1>0$ such that 
		$$		
		\PP{\lambda_{max}(\tilde{p}\tilde{Z}\tilde{Z}^T)>c_1 \text{ or }\lambda_{min}(\tilde{p}\tilde{Z}\tilde{Z}^T)<1/c_1}\le \exp(-C_1n)
		$$
		holds for any $n\times \tilde{p}$ submatrix $\tilde{Z}$ of $Z$ with $c_0n<\tilde{p}\le p$.
		\item
		$Var(Y) = O(1)$, and for some $\kappa \ge0$ and $c_2, c_3>0$,
		$$
		\min_{j\in \calm^*} |\beta_j| \ge \frac{c_2}{n^{\kappa}}, \quad \min_{j\in \calm^*}Cov(\beta_j^{-1}Y, X_j) \ge c_3
		$$ 
		\item
		There are some $\tau \ge 0$ and $c_4>0$ such that 
		$\lambda_{max}(\Sigma) \le c_4 n^{\tau}$.
	\end{enumerate}

\begin{proof}[{\bf Proof of Theorem 1}]
	To prove Theorem \ref{thm:wdist}, we need to use the following lemma, which is from \cite{arratia89two}.
	
	\begin{lemma}
	Let $I$ be an index set and $\{\Ba, \ainI\}$ be a set of subsets of $I$, that is, $\Ba\subset I$ for each $\alpha\in I$. Let also $\{\ea, \ainI \}$ be random variables. For a given $t\in R$, set $\lambda = \sum_{\ainI}\PP{\ea>t}$. Then
	\begin{equation*}
	|\PP{\max_{\ainI}\ea<t}-e^{-\lambda}|\le
	(1\wedge \lambda^{-1})(b_1+b_2+b_3)
	\end{equation*}
	where 
	$b_1 = \sum_{\ainI}\sum_{\beta\in\Ba}\PP{\ea> t}\PP{\eta_{\beta}>t}$, 
	$b_2 = \sum_{\ainI}\sum_{\alpha\ne\beta\in\Ba}\PP{\ea>t, \eta_{\beta}>t}$ and  
	$b_3  = \sum_{\ainI}E|\PP{\ea>t|\sigma(\eta_{\beta}, \beta\notin\Ba)}-\PP{\ea>t}|$,
	and $\sigma(\eta_{\beta}, \beta\notin\Ba)$ is the $\sigma$-algebra generated by $\{\eta_{\beta}, \beta\notin\Ba\}$. In particular, if $\ea$ is independent of $\{\eta_{\beta}, \beta\notin\Ba\}$ for each $\alpha$, then $b_3$=0.
	\end{lemma}
	In our proof, we take $I = \{(i,j); 1\le i\le j\le p \}.$ Let $\alpha = (i,j)\in I, $ we define $\Ba = \{(k,l)\in I;$ one of $k$ and $l = i$ or $j$, but $(k,l)\ne\alpha\}$, and $\Aa = A_{ij} = \{|\rij|^2\ge t\}$, where $\rij = |\widehat{\text{Corr}}(X_i, X_j)|$. Let $\Wpn = \max_{1\le i<j\le p}|\rij|$, by the Chen-Stein method (in particular, Lemma 6.2 in Cai and Jiang (2011)),
	
	\begin{equation}\label{eq:csineq}
	|P(\Wpn^2 \le t)-e^{-\lambn}|\le b_1 + b_2,
	\end{equation}
	where 
$	\lambn = \sum_{\ainI}P(\Aa)= \frac{p(p-1)}{2}P(A_{12})$,
	and
	$b_1 = \sum_{\ainI}\sum_{\beta\in\Ba}P(\Aa)P(A_{\beta})$, 
	$b_2  =  \sum_{\ainI}\sum_{\alpha\ne\beta\in\Ba}P(\Aa A_{\beta})$.
	
	Moreover, we have
$	b_1\le 2p^3P(A_{12})^2 \mbox{ and } b_2\le 2p^3P(A_{12}A_{13})$.
	
	Since $X_1, \cdots, X_p$ are independent, $A_{12}$ and $A_{13}$ are also independent with equal probability. Therefore we have
	$
	b_1\vee b_2\le 2p^3P(A_{12})^2
	$.
	
	On the other hand, $|\rij|^2\sim B(\frac{1}{2}, \frac{n-2}{2})$. Take $t^* = \aapn+\bbpn x~(x\le\frac{n-2}{2})$, where $\aapn = 1-\nnpn\ccpn, \bbpn = \frac{2}{n-2}\nnpn\ccpn$, and $\ccpn = \big(\frac{n-2}{2}\Betan\sqrt{1-\nnpn}\big)^{2/(n-2)}$. Then
	\begin{equation}\label{eq:prho}
	\begin{split}
	P(A^*_{12}) & = \frac{2(1-t^*)^{(n-2)/2}}{B(\frac{1}{2}, \frac{n-2}{2})(n-2)\sqrt{t^*}}(1+O(\frac{1}{\log(p)})).  \\
	& = p^{-2}\big(1-\frac{2}{n-2}x\big)^{\frac{n-2}{2}}
	\sqrt{\frac{1-p^{-4/(n-2)}}{\aapn}}
	\big(1+(\frac{\bbpn}{\aapn}x)\big)^{-1/2}\big(1+O(\log^{-1}(p))\big). \\
	& = p^{-2}\big(1-\frac{2}{n-2}x\big)^{\frac{n-2}{2}}\big(1+O(\frac{\log\log(p)}{\log(p)})\big)\big(1+O(\log^{-1}(p))\big)^2 \\
	& = p^{-2}\big(1-\frac{2}{n-2}x\big)^{\frac{n-2}{2}}\big(1+O(\frac{\log\log(p)}{\log(p)})\big)
	\end{split}
	\end{equation}

	Therefore, uniformly for any $n\ge 3$,
	$b_1\vee b_2 = O(1/p)$, and 
	$\lim_{p\rightarrow\infty} \lambn = \frac{1}{2}\big(1-\frac{2}{n-2}x\big)^{\frac{n-2}{2}}$
	
	Then it follows from \eqref{eq:csineq} that uniformly for any $n\ge 3$ and $x\le \frac{n-2}{2}$,
	\begin{equation}\label{eq:extrmd}
	\lim_{p\rightarrow\infty}|P(\Wpn^2\le t^*) - \exp\big\{-\frac{1}{2}\big(1-\frac{2}{n-2}x\big)^{\frac{n-2}{2}}\big\}|=0.
	\end{equation}
	
	When $x\ge \frac{n-2}{2}$, $t^* = 1 + (\frac{2}{n-2}x -1) p^{-4/(n-2)}\ccpn\ge 1$. Therefore, uniformly for any $n\ge 3$,
	\begin{equation}\label{eq:extrm2}
	\lim_{p\rightarrow \infty}P(\Wpn\le t^*)=1
	\end{equation}
	
	Combining \eqref{eq:extrmd} and \eqref{eq:extrm2} we have uniformly for any $n\ge 3$,
	\begin{equation}
	\lim_{p\rightarrow \infty}|P(\Wpn\le t^*) - I(x\le \frac{n-2}{2}) \exp\big\{-\frac{1}{2}\big(1-\frac{2}{n-2}x\big)^{\frac{n-2}{2}}\big\}-I(x>\frac{n-2}{2})|=0.
	\end{equation}
	
	Or equivalently,
	\begin{equation}
	\lim_{p\rightarrow \infty}|P(\frac{\Wpn^2-\aapn}{\bbpn}\le x) - I(x\le \frac{n-2}{2}) \exp\big\{-\frac{1}{2}\big(1-\frac{2}{n-2}x\big)^{\frac{n-2}{2}}\big\}-I(x>\frac{n-2}{2})|=0.
	\end{equation}
\end{proof}

\begin{proof}[{\bf Proof of Theorem 3}]
	
	Let event $A = \{R^2_{ij}\le 1-p^{-(4+\delta)/(n-3)} \text{ for all }i,j \in\calm\backslash\calm^* \}$,  event $B = \{\hrij \le f(n,p,\alpha) \text{ for }i\in \calm^*, j\in  \calm\backslash\calm^*\}$ where $\hrij = |\hcorr(X_i, X_j)|$,  $f(n,p,\alpha)$ is the screening threshold for pairwise correlation screening. Then $A$ implies that no pairs of unimportant variables passed the R squares screening. $B$ implies that important and unimportant variables can not be too highly correlated.
	
	By the definition of $\calc$, we have 
	
	\begin{equation}
	P(\calc \cap \calm\subset \calm^*) \ge P(A\cap B)\ge P(A)+P(B)-1.
	\end{equation}
	
	For the event $A$, we have 
	\vspace{-.8cm}
	\begin{align*}
	P(A)& = 1- P(\bigcup_{i\ne j\in \calm\backslash\calm^*}R_{ij}^2\ge 1-p^{-(4+\delta)/(n-3)}) \ge 1-\sum_{i\ne j\in  \calm\backslash\calm^*}P(R_{ij}^2\ge 1-p^{-(4+\delta)/(n-3)})\\
	& = 1-(n/\log(n))^2P(\text{Beta}(1, \frac{n-3}{2})\ge 1-p^{-(4+\delta)/(n-3)})\\
	& = 1-(n/\log(n))^2 p^{-(4+\delta)/2}
	\end{align*}
	
	Under the assumption $(B1)$, $(n/\log(n))^2 p^{-(4+\delta)/2}\rightarrow 0$ as $n\rightarrow \infty$. Therefore we have $P(A)\rightarrow 1$.
	
	Next we show that $P(B)\rightarrow 1$ as $n\rightarrow \infty$. We have
	\vspace{-.8cm}
	\begin{align*}
	P(B) &= 1- p(\bigcup_{i\in\calm^*, j\in\calm\backslash\calm^*} \hrij \ge f(n,p, \alpha)) \ge 1- \sum_{i\in\calm^*, j\in\calm\backslash\calm^*}P(\hrij \ge f(n,p, \alpha))\\
	&= 1- (n/\log(n))^2 \PP{\hrij \ge \max\{ \aapn +\bbpn F_n(\alpha), \eta\}} \\
	& = 1- (n/\log(n))^2\PP{\hrij \ge \detnp},
	\end{align*}
	where $F_n(\alpha)$ is the $100(1-\alpha)$ quantile of the limiting cumulative distribution function of the maximal pairwise correlation statistic, and we denote $\max\{ \aapn +\bbpn F_n(\alpha), \eta\}$ by $\detnp$.

	Note that 
	\vspace{-1cm}
	\begin{align*}
	 &\aapn +\bbpn F_n(\alpha) 	=1- p^{-4/(n-2)}\ccpn (1-\frac{2}{n-2}F_n(\alpha)) \\
	= & 1- p^{-4/(n-2)}\ccpn\{-2\log(1-\alpha)\}^{2/(n-2)} \\
	=&1-\Big(C_{\alpha}p^{-2}\frac{n-2}{2}\Betan\sqrt{1-p^{-4/(n-2)}}\Big)^{\frac{2}{n-2}} \\
	=& 1- 
	O\Big(\frac{C^2_{\alpha}(n-2)(1-p^{-4/(n-2)})}{p^4}\Big)^{\frac{1}{n-2}} \quad \text{for large enough } n \\
	= & 1- O\big(e^{-\frac{\log p}{n} }\big)\quad \text{for large enough } n
	\end{align*}
	
	Let $\rho_{ij}$ be the population correlation coefficient between $X_i$ and $X_j$. Write $z(n) = \frac{1}{2}\log\frac{1+\hrij}{1-\hrij}$, $\xi =\frac{1}{2}\log\frac{1+\rho_{ij}}{1-\rho_{ij}} $.  It has been shown that as $n\rightarrow\infty$,
	$
	n^{1/2}(z(n)-\xi) \rightarrow \caln(0, 1).
	$
	
	We have
	\vspace{-.8cm}
	\begin{equation}\label{eq:tailprob}
	\begin{split}
	 \PP{\hrij \ge \detnp} = &
	 \PP{n^{1/2}(z(n)-\xi) \ge n^{1/2}(\frac{1}{2}\log \frac{1+\detnp}{1-\detnp}-\xi)}	 \\
	= & \PP{Z \ge n^{1/2}(\frac{1}{2}\log \frac{1+\detnp}{1-\detnp}-\xi) + o_n(1)} 
	\le  \frac{e^{-C_{p,n}n}}{\sqrt{2\pi n}C_{p,n}},
	\end{split}
	\end{equation}
	where $C_{p,n} = \frac{1}{2}\log \frac{1+\detnp}{1-\detnp}-\xi$. 
	
	If $\log(p)/n\rightarrow \infty$ as $n\rightarrow \infty$, then
	$\aapn + \bbpn F_n(\alpha)\rightarrow 1$. Therefore $\detnp\rightarrow 1$, which yields $C_{p,n}\rightarrow \infty$. Then the tail probability in \eqref{eq:tailprob} goes to zero as $n\rightarrow\infty$. It follows that $P(B)\rightarrow 1$ as $n\rightarrow \infty$.
	
	If $\log(p)/n\rightarrow \eta_0$ as $n\rightarrow \infty$, then $\detnp\rightarrow \max \{1- e^{-4\eta_0}, \eta\}$. Under assumption $(B2)$ that $\rho_{ij}< \max \{1- e^{-4\eta_0}, \eta\}$, $\limn C_{p,n} = \limn \frac{1}{2}\log \frac{1+\max \{1- e^{-4\eta_0}, \eta\}}{1-\max \{1- e^{-4\eta_0}, \eta\}}-\xi >0$. Again the tail probability in \eqref{eq:tailprob} goes to zero as $n\rightarrow\infty$. It follows that $P(B)\rightarrow 1$ as $n\rightarrow \infty$.
	
	If $\log(p)/n\rightarrow 0$ as $n\rightarrow \infty$, then $\aapn + \bbpn F_n(\alpha)\rightarrow 0$. Hence $\detnp\rightarrow \eta$. Under the assumption $(B2)$, we have $\limn C_{p,n} = \log \frac{1+\eta}{1-\eta} -\xi >0$. Therefore $P(B)\rightarrow 1$ as $n\rightarrow \infty$.

	Given $P(A)\rightarrow 1$ and $P(B)\rightarrow 1$, we have $P(\calc \cap \calm\subset \calm^*)\rightarrow 1$ as $\infn$. 	
\end{proof}

\begin{proof}[{\bf Proof of Theorem 4}]
	It follows from \eqref{eq:zy} directly that
	\begin{equation}\label{eq:irrsub}
	\|(C_{21}^{(2)}-C_{21}^{(1)}(C_{11}^{(11)})^{-1}C_{11}^{(12)})\big(C_{11}^{(22)}-C_{11}^{(21)}(C_{11}^{(11)})^{-1}C_{11}^{(12)}\big)^{-1}\text{sign}(\bbeta_1^{(2)})\|_{\max}\le 1-\xi,
	\end{equation}
	where $\| \cdot \|_{\max}$ denotes the max norm of a matrix.	Based the definition of $\calc$, we have the following element wise inequalities $\|C_{11}^{(12)}\|_{\max}\le ~c_{n,p, \alpha}$, 	$\|C_{11}^{(21)}\|_{\max}\le c_{n,p, \alpha}$. Here $c_{n,p, \alpha}$ is the pairwise correlation screening bound. Since $C_{11}^{(11)}$ is positive definite, there exists an orthogonal matrix $Q$ s.t. $C_{11}^{(11)} = Q\Lambda Q^T$, where $\Lambda$ is a diagonal matrix consists of the eigenvalues of $C_{11}^{(11)}$. By assumption, we have $\lambda_{min}(C_{11}^{(11)})\ge \lambda_0$. Therefore
	$
	\|C_{11}^{(21)}(C_{11}^{(11)})^{-1}C_{11}^{(12)}\|_{\max}\le \lambda^{-1}_0c^2_{n,p, \alpha}s_1^2.
	$
	Under the assumption that $\log(p)/n \rightarrow 0$, $c_{n,p, \alpha} = o_n(1)$. It follows that $\lambda^{-1}_0c^2_{n,p, \alpha}s_1^2 = o_n(1)$.	
	By assumption $(B2)$, $\|C_{21}^{(1)}\|_{\max}\le \eta$. Thus
	$\|C_{21}^{(1)}(C_{11}^{(11)})^{-1}C_{11}^{(12)}\|_{\max}\le~\lambda^{-1}_0\eta c_{n,p, \alpha}s_1^2$,
	then $\|C_{21}^{(1)}(C_{11}^{(11)})^{-1}C_{11}^{(12)}\|_{\max}$ = $o_p(1)$ as $n\rightarrow\infty$. 
	Therefore
	\begin{align*}
	& \|(C_{21}^{(2)}-C_{21}^{(1)}(C_{11}^{(11)})^{-1}C_{11}^{(12)})\big(C_{11}^{(22)}-C_{11}^{(21)}(C_{11}^{(11)})^{-1}C_{11}^{(12)}\big)^{-1}\text{sign}(\bbeta_1^{(2)})-C_{21}^{(2)}(C_{11}^{(22)})^{-1}\text{sign}(\bbeta_1^{(2)})\|_{\max} \\
	= & \|\big( C_{21}^{(2)}(C_{11}^{(22)})^{-1} C_{11}^{(21)}(C_{11}^{(11)})^{-1}C_{11}^{(12)} -C_{21}^{(1)}(C_{11}^{(11)})^{-1}C_{11}^{(12)} \big)
	\big(C_{11}^{(22)}-C_{11}^{(21)}(C_{11}^{(11)})^{-1}C_{11}^{(12)}\big)^{-1}\text{sign}(\bbeta_1^{(2)}) \|_{\max}
	\end{align*}
	Write $A = C_{21}^{(2)}(C_{11}^{(22)})^{-1} C_{11}^{(21)}(C_{11}^{(11)})^{-1}C_{11}^{(12)}, B = C_{21}^{(1)}(C_{11}^{(11)})^{-1}C_{11}^{(12)}, D = C_{11}^{(21)}(C_{11}^{(11)})^{-1}C_{11}^{(12)}$, and $Y =\text{sign}(\bbeta_1^{(2)})$. Then the above term becomes $\|(A - B)(C_{11}^{(22)} - D)^{-1}Y\|_{\max}$. Moreover, we have
	\vspace{-.8cm}
	\begin{align*}
	\|(A - B)(C_{11}^{(22)} - D)^{-1}Y\|_{\max}\le (s-s_1) \|A-B\|_{\max}\|(C_{11}^{(22)} - D)^{-1}Y\|_{\max}.
	\end{align*}
	Since $
	\|A\|_{\max}\le \lambda_0^{-1}(s-s_0)^2 \|C_{21}^{(2)}\|_{\max}\|C_{11}^{(21)}(C_{11}^{(11)})^{-1}C_{11}^{(12)}\|_{\max} \le \lambda^{-2}_0\eta c^2_{n,p, \alpha}s_1^2(s-s_1)^2,
	$	
	$\|B\|_{\max}\le\lambda^{-1}_0\eta c_{n,p, \alpha}s_1^2$, and 
	$$
	\|(C_{11}^{(22)} - D)^{-1}Y\|_{\max} \le (s-s_1)\|(C_{11}^{(22)} - D)^{-1}\|_{\max} 
	\le (s-s_1)(\lambda_0-\lambda^{-1}_0c^2_{n,p, \alpha}s_1^2)^{-1}.
	$$	
	Therefore we have
		\vspace{-.8cm}
	\begin{align*}
	\|(A - B)(C_{11}^{(22)} - D)^{-1}Y\|_{\max} &\le (s-s_0)^2(\lambda^{-2}_0\eta c^2_{n,p, \alpha}s_1^2(s-s_1)^2 + \lambda^{-1}_0\eta c_{n,p, \alpha}s_1^2)(\lambda_0-\lambda^{-1}_0c^2_{n,p, \alpha}s_1^2)^{-1} \\
	& = o_p(1),
	\end{align*}
	as $n\rightarrow \infty$. It follows that
	 $
	C_{21}^{(2)}(C_{11}^{(22)})^{-1}\text{sign}(\bbeta_1^{(2)})<1-\xi/2$ with probability tending to 1 as $n\rightarrow \infty$ which concludes the proof if we take $\delta = \xi/2$.
\end{proof}